\documentstyle[prl,aps,floats,epsf]{revtex}
\begin{document}
\twocolumn[
\hsize\textwidth\columnwidth\hsize\csname@twocolumnfalse\endcsname
\draft    
\begin{center}
To be published in the Canadian Journal of Physics, Boris P. Stoicheff
special issue.
\end{center}
\title{The Pseudogap in  
La$_{\bf {2-x}}$Sr$_{\bf {x}}$CuO$_{\bf {4}}$: A Raman Viewpoint}
\author{J. G. Naeini, J. C. Irwin}
\address{Department of Physics, Simon Fraser University, Burnaby, 
British Columbia, V5A 1S6, Canada}
\author{T. Sasagawa, Y. Togawa, and K. Kishio}
\address{Department of Superconductivity, University of Tokyo, 
Bunkyo-ku, Tokyo 113, Japan}
\maketitle
\begin{abstract} 
We report the results of Raman scattering experiments on single 
crystals of La$_{\rm {2-x}}$Sr$_{{\rm x}}$CuO$_{4}$ (La214) as a function 
of temperature and doping. 
In underdoped compounds low-energy  B$_{1g}$ spectral weight is depleted
in association with the opening of a pseudogap on regions of the Fermi
surface located near ($\pm \pi,0$) and ($0, \pm \pi$).
The magnitude of the depletion increases with decreasing doping, and in the 
most underdoped samples, with decreasing temperature.
The spectral weight that is lost at low-energies ($\omega \leq 800\,cm^{-1}$)
is transferred to the higher energy region normally occupied by 
multi-magnon scattering. From the normal state B$_{2g}$ spectra 
we have determined the scattering rate $\Gamma(\omega$,T) of qausiparticles 
located near the diagonal directions in k-space.
In underdoped compounds, $\Gamma(\omega$,T) is suppressed 
at low temperatures for energies less than 
$E_g(x) \simeq 800\,cm^{-1}$. 
The doping dependence of both the two-magnon scattering and the 
scattering rate suppression suggest that the pseudogap is characterized
by an energy scale E$_g \sim J$, 
where $J$ is the antiferromagnetic super-exchange energy.
Comparison with the results from other techniques provides
a consistent picture of the pseudogap in La214.
\end{abstract} 
]

\section{Introduction}
It is now clear \cite{randeria} that the unusual electronic properties 
of the underdoped high temperature superconductors are strongly influenced 
by the presence of a normal state pseudogap (PG). 
However, the origin of this PG and any relation it might 
have to the occurrence of superconductivity, remain controversial issues.  
In an attempt to resolve these questions many recent experiments have 
attempted to determine the energy scale E$_g$, doping dependence, and 
symmetry associated with the PG.  The experimental situation remains unclear, 
however, in that Raman \cite{chuck,jafar,chen,nemet,opel,quilty}, 
far infrared 
reflectivity (FIR) \cite{tatiana,puchkov,wang}, specific heat \cite{loram} and 
some recent \cite{ino} angle resolved photoemission spectroscopy 
(ARPES) experiments yield E$_g \sim J$, where $J$ is the 
antiferromagnetic (AFM) exchange energy, 
while tunneling \cite{renner} and earlier ARPES experiments 
\cite{marshall,ding} on Bi$_2$Sr$_2$CaCu$_2$O$_{\rm z}$ (Bi2212)
found E$_g \sim \Delta$, the superconducting gap energy.  
The results of ARPES experiments \cite{marshall,ding} also 
suggest that the PG has conventional d-wave symmetry while Raman results 
\cite{jafar,nemet}
indicate that the gap is more localized to regions of the Fermi surface 
(FS) located near the axes in reciprocal space.  Finally FIR 
\cite{tatiana,puchkov}, ARPES
\cite{marshall,ding}, and NMR \cite{bp,tallon} experiments
indicate that at certain doping levels the PG is characterized by 
an onset temperature which is not well defined 
in specific heat \cite{loram,tallon} and some Raman \cite{chuck,jafar} 
measurements.  

In an attempt to reconcile some of the apparent conflicts mentioned above 
we have carried out a systematic 
Raman investigation of the dependence of the PG parameters on doping 
and temperature in La$_{\rm {2-x}}$Sr$_{{\rm x}}$CuO$_{4}$ (La214).   
This compound is an excellent material for these studies for several reasons. 
It has a relatively simple structure with a single CuO$_2$ plane in the 
primitive cell. Thus the results are not influenced by structural 
complications such as those introduced by the  chains in 
YBa$_2$Cu$_3$O$_y$ (Y123), or the 
structural modulations in Bi2212. 
Furthermore, the hole concentration is 
determined simply by the Sr concentration if oxygen stoichiometry is 
maintained. As a result, one can obtain \cite{kimura} high quality, 
well characterized single crystals of La214 that enable one to study the 
systematic evolution of the electronic properties throughout the 
complete doping range.  
The samples studied in this work were grown by a traveling floating-zone 
method \cite{kimura} and were carefully characterized using x-ray diffraction, 
transport and susceptibility measurements \cite{kimura}. The physical 
parameters of the La214 samples are summarized in (Table I). 
\begin{table}[htb]
\centering
\begin{tabular}{|c|c|c|}
La$_{\rm {2-x}}$Sr$_{{\rm x}}$CuO$_{4}$&Sr content (x)&T$_c$ 
(K)\\ \hline 
Underdoped&0.08&16\\ \hline
Underdoped&0.11&27\\ \hline 
Underdoped&0.13&35\\ \hline  
Optim-doped&0.17&37\\ \hline
Overdoped&0.19&32\\ \hline 
Overdoped&0.22&30\\  
\end{tabular}
\vspace{0.1in}
\caption{The physical parameters characterizing the samples studied in 
this paper. T$_c$ was determined magnetically and 
x is the nominal Sr concentration (Ref. 18).}
\label{tab:xTc}
\end{table}

We have previously investigated \cite{jafar,chen} the 
B$_{1g}$ and B$_{2g}$ Raman spectra of La214 for doping levels  
$0.13 \leq x \leq 0.22$, and found that in underdoped materials 
there is a significant depletion of B$_{1g}$  
spectral weight at low-energies ($\omega \leq 800\,cm^{-1}$). 
Furthermore, the strength of this depletion increases rapidly as the doping 
level is decreased below optimum.  
In the La214 samples studied previously,  
and in other cuprates \cite{chen,nemet}, the strength of the depletion
was found to be approximately independent of temperature for $T \leq 300\,$K.
Finally, in optimally doped compounds the B$_{1g}$ spectra 
undergo \cite{xiaoke}  a strong renormalization   that results in
the formation of a 2$\Delta$ peak for T$\,\leq\,$T$_c$.
In all the underdoped cuprates, however, the B$_{1g}$ spectrum is unaffected
by the superconducting transition at T$_c$.  
In contrast to the significant doping induced 
changes that occur in the B$_{1g}$ spectra,
the strength of the B$_{2g}$ spectra
appear to be \cite{chuck,jafar,chen} approximately independent of doping.
In addition, a superconductivity induced renormalization \cite{jafar} is 
observed in the B$_{2g}$ channel for all doping levels.

In this paper we have extended our range of investigation of La214 
to lower doping levels ($x = 0.08$ and $x = 0.11$) and to higher energies
($\omega \leq 4000\,cm^{-1}$). Spectra have been obtained
in both B$_{1g}$ and B$_{2g}$  channels that corroborate previously observed
trends \cite{jafar,chen} in the doping dependence of 
the B$_{1g}$ spectral weight depletion, and the relative 
immunity of the B$_{2g}$ spectra to changes in doping.
In addition, in the most underdoped samples ($x = 0.08$ and $x = 0.11$) 
the B$_{1g}$ spectral weight at low energies is further
depleted as the temperature is reduced from 300K to 50K. The spectra also
show that this spectral weight appears to be 
transferred from the low-energies region  
to the higher energy ($\omega \,>\, 1500\,cm^{-1}$) spectral region
normally occupied by multi-magnon scattering.
The observed temperature dependence of the scattered intensity in the 
B$_{1g}$ channel is found to be correlated with a depression of the 
scattering rate $\Gamma(\omega$,T) in the B$_{2g}$ channel, 
similar to that derived from measurements \cite{tatiana,puchkov,wang} of 
the infrared conductivity.
The suppression of $\Gamma(\omega$,T) is characterized by an energy scale 
$\omega_{\Gamma}(x) \sim 800\,cm^{-1}$ which is the same \cite{chuck,jafar} 
as the energy scale 
associated with the doping induced depletion of spectral weight.
The results are consistent with those obtained using other techniques and
suggest that the PG is magnetic in origin.

\section{Results and Discussion}
\subsection{Spectral Weight Depletion}
\begin{figure}[htb]
\centerline{\epsfxsize=3.2 in \epsffile{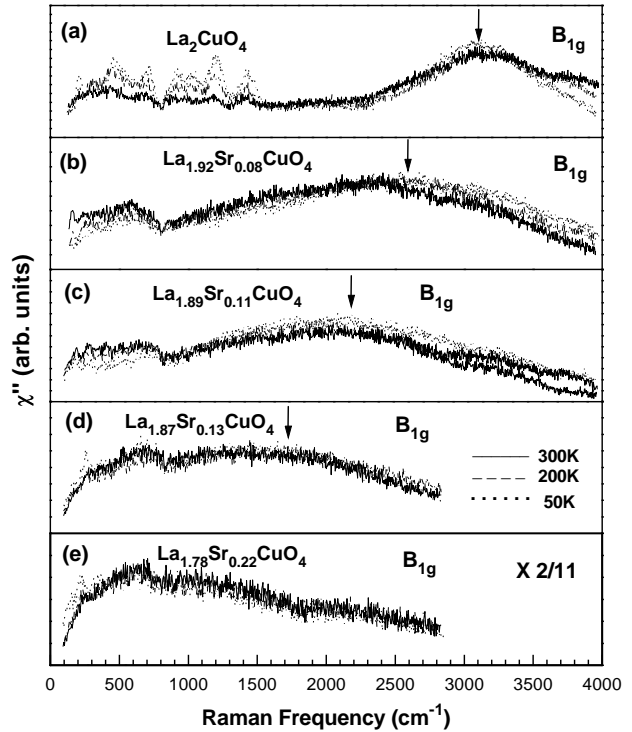}}
\vspace{0.1in}
\caption{The B$_{1g}$ Raman response functions 
$\chi^{\prime\prime}(\omega,$T) measured at temperatures between
50K and 300K in La214(x) for (a) x = 0; (b) x = 0.08; (c) x = 0.11;
(d) x = 0.13; and (e) x = 0.22.}
\label{fig1}
\end{figure}
In this paper all the as-recorded spectra have been divided by the Bose-Einstein factor to obtain 
the relevant Raman response functions.
The B$_{1g}$ Raman response function of La214 is shown in Fig. 1 for 
different doping levels. It should be noted that all the 
spectra shown in Fig. 1 are plotted on the same scale (after dividing the 
x = 0.22 spectrum by 5.5). This means, for example, that for the underdoped 
(x = 0.13) crystals, the integrated spectral weight in the B$_{1g}$ channel 
for $0 \leq \omega \leq 800\,cm^{-1}$,  is about a factor of 6 times smaller 
than that for the overdoped (x = 0.22) crystal.  
Since the B$_{1g}$ spectra are dominated by scattering from excitations on 
regions of the FS located near the k$_x$ and k$_y$ axes \cite{chen,xiaoke},
the results imply that underdoping leads to a significant depletion of 
low-energy spectral weight from these same regions of the FS
[hereafter designated as the ($\pi,0$) regions].  We 
attribute \cite{chuck,jafar,chen} this depletion of spectral weight to 
the presence of a PG which is approximately independent of temperature for 
doping levels $x \geq 0.13$ and temperatures $T \leq\,$300K.  

\begin{figure}[htb]
\centerline{\epsfxsize=3.1 in \epsffile{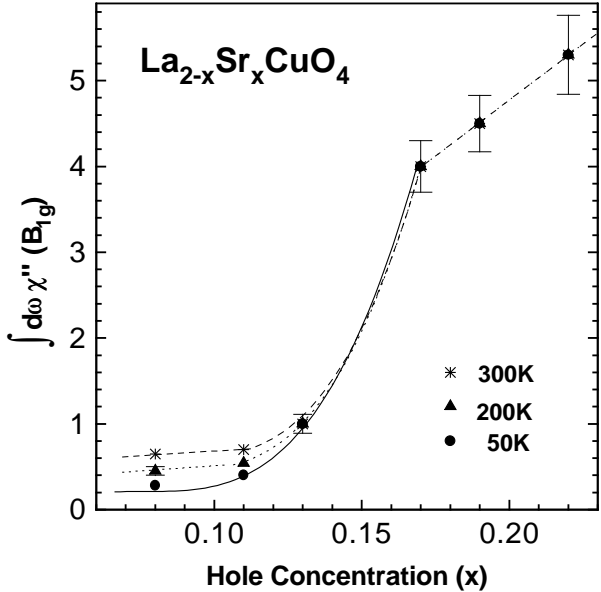}}
\vspace{0.1in}
\caption{The integrated low-energy ($0 \leq \omega \leq 800\,cm^{-1}$)
B$_{1g}$ spectral weight  plotted as a function of hole concentration
at temperatures between 50K and 300K in La214. 
The data for x = 0.17 and 0.19 crystals are the same as those obtained at
$T \, < \, T_c$  (Ref. 3) and have not been shown in Fig. 1 for clarity.
The solid line is the result of a calculation (see text) while 
the dashed and dotted lines serve only as guides to the eye.}
\label{fig2}
\end{figure}
The doping dependence of the low energy spectral weight is summarized
in figure 2.  As is evident from Fig. 2 the strength of the depletion
increases rapidly with decreasing doping.  Alternatively one can say
that the PG removes spectral weight from regions of the Fermi surface
located near ($\pi, 0$) and as we underdope the size of the affected
regions increases significantly.  The solid line in Fig. 2
was calculated \cite{chuck,jafar} using a simple tight binding band
structure and assuming that the depleted arc length of
the Fermi surface increases with decreasing doping.
The qualitative agreement between experiment and calculation
supports this effective fragmentation of the Fermi surface.
Furthermore, the additional reduction in spectral weight
that occurs at low T, suggests that decreasing temperature
leads to a further shrinking of the active Fermi surface
area, as observed in a previous ARPES experiment \cite{norman}.

As the doping level of La214 is decreased below optimum a broad peak appears 
in the B$_{1g}$ spectra at higher energies (Fig. 1).  
This peak is attributed to 2-magnon scattering and it grows in strength as 
the doping level is decreased.  For $x \leq 0.11$ it becomes the dominant 
feature of the B$_{1g}$ spectra.   
Underdoping thus leads to an effective transfer of spectral weight from low 
energies to the higher frequency range occupied by the 2-magnon features.  
This observation is consistent with the suggestion \cite{chuck,jafar}  
that the depletion of low energy spectral weight is associated with the presence
of short range magnetic correlations. Such correlations are assumed to grow 
in strength as the 
doping level is decreased \cite{bp,pp,tom}, thus removing an 
increasing amount of spectral weight (Fig. 2) from the FS and leading to 
an effective fragmentation \cite{jafar} of the FS.  A more critical  
consideration of the spectra, however, indicates that the spectral weight
gained in the 2-magnon frequency region is greater than the amount lost
below 800 cm$^{-1}$, both as a function of doping and as a function of
temperature.  Although this observation may be somewhat surprising one
\begin{figure}[htb]
\centerline{\epsfxsize=3.25 in \epsffile{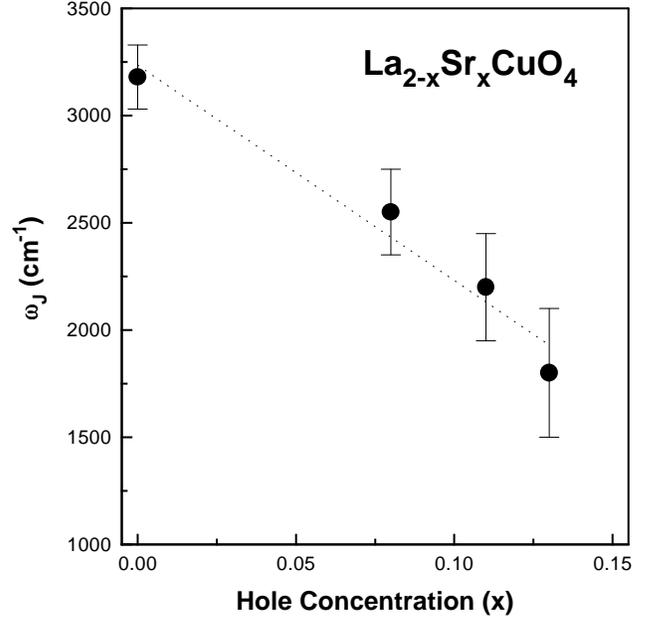}}
\vspace{0.1in}
\caption{The variation of the two-magnon frequency 
($\omega_J$) with doping (x) in La214.
The dotted line serves only as a guide to the eye.}
\label{fig3}
\end{figure}
must remember that the Raman intensity is proportional to a generalized 
density-density correlation function and thus one would not expect sum 
rules to apply rigorously \cite{opel1}.  In addition one must also note
that the Raman continua extend to very high energies ($\sim$2eV), which are 
outside the range of our spectrometer.  It is possible, 
and perhaps even expected, that spectral weight will also be transferred 
from the region $\omega >$ 0.5eV to the 2-magnon region, as the doping 
level or temperature are reduced. Thus it is perhaps not too surprising 
that the observed enhancement in the 2-magnon region appears to exceed 
the low energy depletion.   

In Raman experiments one-magnon scattering has not been observed 
but the magnetic excitations can be probed via the two-magnon scattering 
which is peaked \cite{sugai,blum,rub} at $\omega_J \sim 3J$ in the undoped 
cuprates (Fig. 1).  
As the doping level is increased, and the AFM correlation length decreases, 
this feature broadens, weakens and softens in frequency.  The peak 
frequencies associated with these 2-magnon features (which are designated
$\omega_J$), have been estimated as shown in Fig. 1.  The resultant  
values of $\omega_J$ are plotted as a function of doping in Fig. 3. 
If $\omega_J$ is assumed to be a measure of 
the strength of the short range AFM correlations the energy of the PG is 
then given by E$_g \sim \omega_J / 3 \sim  J\, (1 - x / 0.3)$ 
for $0 \leq x \leq 0.13$.  Such a linear 
variation for E$_g$ was initially proposed by Loram {\em et al.}  
\cite{loram} from the results of specific heat measurements.  We must 
emphasize however that the identification of the peak positions 
($\omega_J$) is highly uncertain and thus the linear relation shown
in Fig. 3 must be considered to be very tentative.
It should also be  noted that the spectra shown in Fig. 1
have not been corrected for variations in the optical constants or for 
spectrometer response. 
Our spectrometer response decreases in the red and 
correction might push the peaks in Figs. 1(a-d), 
and hence the values of $\omega_J$ shown in Fig.1, to higher energies. 
Such corrections would not, however, alter the observed trend of
$\omega_J$ to lower values as the doping level is increased. 

\subsection{Temperature Dependence}
In the most underdoped samples the low-energy  
B$_{1g}$ spectral weight  also decreases with decreasing  
temperature (Fig. 1).  The integrated loss over this region 
($\omega \leq 800\,cm^{-1}$), as a function of temperature, 
is summarized in Fig. 2.  It is clear that the temperature dependent 
depletion is smaller in magnitude, and appears to be superimposed 
on the doping induced depletion.  The depletion induced by underdoping 
sets in at a temperature that is outside our range of observation 
(T $\,>$ 300K).  The results thus imply that there is an upper crossover 
temperature at which strong PG behavior appears and a lower crossover 
temperature T$^*$ below which a further contribution to the PG occurs.  
These results suggest the presence of 
two steps, or perhaps  two mechanisms, associated with 
the depletion of low-energy spectral weight in underdoped crystals.  
The values obtained for T$^*$ from Fig. 1 for the 
x = 0.08, 0.11, 0.13 crystals are in qualitative agreement with 
estimates obtained for T$^*$ from  FIR \cite{tatiana} and NMR
\cite{bp} measurements in La214. Our results thus suggest that the PG is 
reflected in NMR \cite{bp,tallon} and FIR \cite{tatiana,puchkov,wang} 
measurements only in terms of the weaker 
temperature dependent depletion that occurs at T$^*$.  
The results also provide 
a reconciliation with specific heat measurements \cite{loram}, in which no 
clear evidence for T$^*$ is obtained.  Since these 
results are determined by an average of the spectral weight around 
the FS they will be most strongly influenced by the doping induced depletion. 

\begin{figure}[htb]
\centerline{\epsfxsize=3.1 in \epsffile{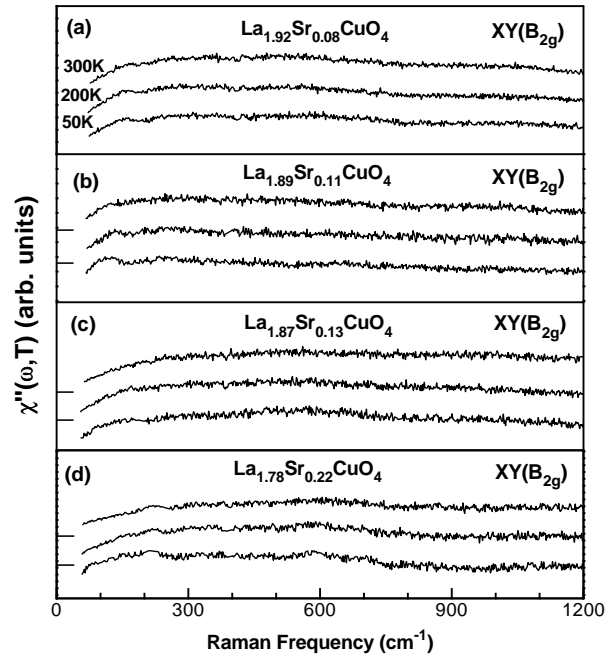}}
\vspace{0.1in}
\caption{The imaginary part of B$_{2g}$ response functions 
$\chi^{\prime\prime}(\omega,$T) measured 
at temperatures between 50K and 300K in La214(x) 
for (a) x = 0.08; (b) x = 0.11; (c) x = 0.13; and (d) x = 0.22. 
The spectra are all plotted on same scale and offset vertically for clarity.}
\label{fig4}
\end{figure}
The presence of short range AFM correlations can give rise to an 
anisotropic scattering rate \cite{pp,tom} that results in the existence of 
``hot spots'' near ($\pi,0$) and ``cold spots'' on regions of the FS near 
($\pm \pi/2,\pm \pi/2$).  The depletion of the B$_{1g}$ spectra (Fig. 2) is 
consistent with the presence of ``hot spots'' and the properties of the 
``cold quasiparticles'' should be reflected in the B$_{2g}$ spectra.  
These spectra (Fig. 4) are relatively independent of doping 
which is consistent with the presence of ``cold spots'' near ($\pm \pi/2,\pm \pi/2$).   
This is somewhat puzzling however in that in infrared experiments
 evidence for the PG has been obtained from the 
frequency and temperature dependence of the scattering rate 
$1/\tau^*(\omega,T)$ that is derived from measurements of the optical
conductivity \cite{tatiana,puchkov}. In underdoped compounds the FIR response should be
dominated by the ``cold'' quasiparticles \cite{jafar,nemet,im}. 
Thus this behavior should also be
reflected in the B$_{2g}$ Raman response. However the B$_{2g}$ spectra
we have obtained (Fig. 4) do not 
exhibit any obvious variation with either doping or temperature. 
The latter result is also in contrast to Raman
data \cite{nemet} obtained from underdoped Bi2212, where the 
B$_{2g}$ intensity was found to decrease by about 10\% when T was reduced
from 250K to 20K.

\section{Comparison with Infrared Spectra}
To investigate the temperature dependence of the B$_{2g}$ spectra 
more carefully  we examine the variation 
of the  scattering rate. We  assume that
the B$_{2g}$ response function can be described approximately by 
\cite{einzel}$\,:$ 
\begin{eqnarray}
\chi(\omega) \propto \frac{i/\tau^*}
{\omega + i/\tau^*}\,. 
\label{metal}
\end{eqnarray}
Qualitative agreement with the experimental spectra of Bi2212 
is obtained \cite{vr} if the scattering rate $1/\tau^*$ is assumed to vary 
with both frequency and temperature. 
To proceed we use an  extended Drude model \cite{puchkov} in which 
 $1/\tau^*$ is given by$\,:$  
\begin{eqnarray}
1/\tau^* \rightarrow \Gamma(\omega,T) - i \omega \lambda(\omega,T)\,, 
\label{IR}
\end{eqnarray}
where  $\Gamma(\omega,T)$ is the scattering rate, and the function 
$\omega \lambda(\omega,T)$ is introduced to preserve causality
\cite{puchkov}. Using this form of $1/\tau^*$,
we arrive at an expression for the {\em extended Drude} response$\,:$ 
\begin{figure}[htb]
\centerline{\epsfxsize=3.1 in \epsffile{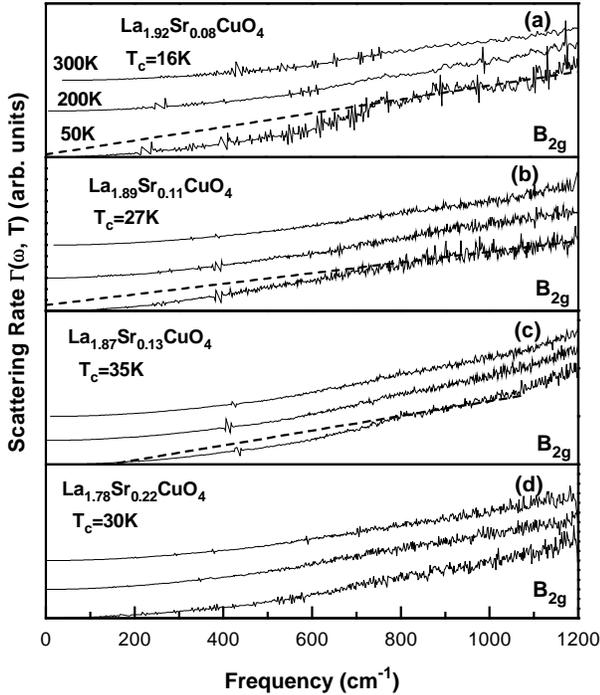}}
\vspace{0.1in}
\caption{The scattering rate $\Gamma(\omega,$T) determined at 
temperatures between 50K and 300K from the B$_{2g}$ Raman spectra of
La214(x) for (a) x = 0.08; (b) x = 0.11; (c) x = 0.13; and (d) x = 0.22.}
\label{fig5}
\end{figure}
\begin{eqnarray}
\chi(\omega,T) \propto \frac{1}
{1 + \omega / [i\Gamma(\omega,T) + \omega \lambda(\omega,T)]}\,. 
\label{exdrud}
\end{eqnarray}
Therefore we can obtain the scattering rate in terms of the 
response function$\,:$
\begin{eqnarray}
\Gamma(\omega,T) 
\propto \frac{\omega \chi^{\prime\prime}(\omega,T)}
{[1 - \chi^{\prime}(\omega,T)]^2 + \chi^{\prime\prime2}(\omega,T)}\,, 
\label{rate}
\end{eqnarray}
where the real part of the response function $\chi^{\prime}(\omega,T)$ can 
 be obtained from $\chi^{\prime\prime}(\omega,T)$ using a
Kramers-Kronig transformation$\,:$
\begin{eqnarray}
\chi^{\prime}(\omega,T) = \frac{2}{\pi} \,P
\int_{0}^{\infty}
\frac{\omega^{\prime}\chi^{\prime\prime}(\omega^{\prime},T)}
{\omega^{\prime \,2} - \omega^2}\,d\omega^{\prime}\,. 
\label{real}
\end{eqnarray}

We have used Eq. (\ref{real}) to evaluate $\chi^{\prime}(\omega,$T) 
and thus determine the scattering rates $\Gamma(\omega,$T) that are
shown in Fig. 5. To carry out the Kramers-Kronig integration,  
$\chi^{\prime\prime}(\omega,$T) is assumed to extrapolate linearly 
to zero at low-frequencies \cite{hackl}. To ensure convergence
$\chi^{\prime\prime}(\omega,$T) must also be truncated
at high-frequencies. Since the results shown in Fig. 4 suggest a linear 
decrease at high-frequencies we have 
extrapolated $\chi^{\prime\prime}$(B$_{2g}$)
linearly to zero at $\Omega \approx 10000 cm^{-1}$. 
It should be noted however, that the results 
are not sensitive to the values of $\Omega$, and results similar to those 
shown in Fig. 5 can also be obtained using either an exponential decay or 
$1/\omega$ extrapolation at high-frequencies. 

As shown in Figs. 5(a-c) the scattering rate $\Gamma(\omega,$T)  
is suppressed (below a linear dependence) at 50K 
in the $x$ = 0.08, 0.11, and 0.13 crystals.
Such a depression could not be observed for $\Gamma(\omega,$T)
obtained at doping levels larger than $x$ = 0.13 (see Fig. 3d, for example).
The frequency $\omega_{\Gamma}$, below which $\Gamma(\omega,$T) is depressed
(indicated by arrows in Fig.5), is approximately equals $800\,cm^{-1}$.
Then, if the frequency $\omega_{\Gamma}$ can be taken \cite{tatiana} as 
a measure of the  PG energy, we can write 
$E_g(x) \simeq 800\,cm^{-1}$. It is interesting to note that the energy scale
of the depletion observed in the B$_{1g}$ channel is correlated 
with the suppression of $\Gamma(\omega,$T) in the B$_{2g}$ channel. 
The results shown in Fig. 5
are consistent with the depression of low-energy scattering rates
that have been obtained from FIR spectra \cite{tatiana,puchkov,wang}.
This observation is also consistent with the suggestion \cite{jafar,nemet,im}
that FIR spectra are determined by the properties of B$_{2g}$ or 
``cold'' quasiparticles.

\section{Conclusions} 
We have carried out Raman scattering investigation of the 
pseudogap in La$_{\rm {2-x}}$Sr$_{{\rm x}}$CuO$_{4}$ as a function of 
both doping and temperature. In underdoped compounds the presence of the 
pseudogap results in a depletion of spectral weight from regions of the Fermi 
surface located near ($\pi,0$), which in turn is manifested by a loss
of spectral weight in the B$_{1g}$ Raman spectra. Reducing the temperature
from 300K to 50K leads to a further reduction of the low-energy
spectral weight which is clearly observed only in the most underdoped
samples. Furthermore, our spectra indicate that spectral weight lost
from the low-energy region is transferred to higher frequencies,
namely to the multimagnon region characterized by $\omega_J \sim 3J$.
We have also shown that the scattering rate depression that is 
used to characterize the pseudogap in infrared measurements can be obtained
from the B$_{2g}$ Raman spectra which implies that the infrared
spectra are dominated by contributions from ``cold'' quasiparticles.
It is interesting to note that the Raman, infrared
\cite{tatiana}, specific heat \cite{loram}, 
and ARPES measurements \cite{ino} on La214 all suggest an energy E$_g \sim J$. 
Thus it appears that an experimental consensus for the pseudogap energy 
scale is emerging, at least in La214. 
Finally  the energy scale and doping dependence that we have observed 
appear to be consistent with a pseudogap associated with short range 
antiferromagnetic correlations.

\begin{center}
{\bf ACKNOWLEDGEMENTS}
\end{center}

The financial support of the Natural Sciences and Engineering Research 
Council of Canada, 
and Core Research for Evolutional Science and 
Technology of Japan Science and Technology 
Corporation are gratefully acknowledged.

\end{document}